\documentclass[twocolumn,runningheads,natbib]{svjour2}
\smartqed  % flush right qed marks, e.g. at end of proof
\usepackage{graphicx}
\newcommand{\apj}[2]{\mbox{ Ap.J.,\ { {\bf #1}},{ #2}}}

\newcommand{\aanda}[2]{\mbox{ A.\&A., { {\bf #1}},{ #2}}}

\newcommand{\mnras}[2]{\mbox{ M.N.R.A.S.,{ {\bf #1}},{ #2}}}

\newcommand{\nat}[2]{\mbox{ Nature, { {\bf #1}},{ #2}}}

\newcommand{\apjl}[2]{\mbox{ Ap.J.Letters, { {\bf#1}},{ #2}}}
\journalname{Astrophysics and Space Science}
\begin{document}

\title{Newborn Magnetars as sources of Gravitational Radiation: constraints
from High Energy observations of Magnetar Candidates} 
\titlerunning{GWs from newborn magnetars}   
% if too long for running head

\author{S. Dall'Osso              
        \and  
        L. Stella
        }

\authorrunning{Dall'Osso et al.}
% if too long for running head

\institute{S. Dall'Osso \at{INAF-Osservatorio Astronomico di Roma\\ via 
di Frascati 33, 00040, Monteporzio Catone (Roma)\\
              Tel.: +39 06 94 28 64 37}
              \email{dallosso@mporzio.astro.it}       
           \and 
L. Stella \at{INAF-Osservatorio Astronomico di Roma\\  via 
di Frascati 33, 00040  Monteporzio Catone (Roma)\\
              Tel.: +39 06 94 28 64 36}
%              Fax: +123-45-678910\\
              \email{stella@mporzio.astro.it}               
}
\date{Received: date / Accepted: date}
\maketitle

\begin{abstract}
Two classes of high-energy sources, the Soft Gamma Repeaters and the 
Anomalous X-ray Pulsars  are believed to contain slowly spinning ``magnetars'',
{\it i.e.} neutron stars the emission of which derives from the release of 
energy from their extremely strong magnetic fields ($>10^{15}$~G).
The enormous energy liberated in the 2004 December 27 giant flare from 
SGR~1806-20 ($\sim 5 \times 10^{46}$~erg), together with the likely recurrence 
time of such events, points to an internal magnetic field strength of 
$ \geq 10^{16}$~G. Such strong fields are expected to be generated by a 
coherent $\alpha-\Omega$ dynamo in the early seconds after the Neutron Star
(NS) formation, if its spin period is of a few milliseconds at most. 
A substantial deformation of the NS is caused by such fields and, provided the 
deformation axis is offset from the spin axis, a newborn millisecond-spinning 
magnetar would thus radiate for a few days a strong gravitational wave signal 
the frequency of which ($\sim 0.5-2$~kHz range) decreases in time. 
This signal could be detected with Advanced LIGO-class detectors up to the 
distance of the Virgo cluster, where  $\geq 1$~yr$^{-1}$ magnetars are 
expected to form. 
%%%%%%
%%%%%%
Recent X-ray observations revealed that SNRs around magnetar candidates do not 
appear to have received a larger energy input than in standard SNRs 
\citep{Vink}. This is at variance with what would be expected if the spin 
energy of the young, millisecond NS were radiated away as electromagnetic 
radiation andd/or relativistic particle winds. In fact, such energy would be
transferred quickly and efficiently to the expanding gas shell. This may thus
suggest that magnetars did not form with the expected very fast initial spin.
We show here that these findings can be reconciled with the idea of magnetars 
being formed with fast spins, if most of their initial spin energy is radiated 
thorugh GWs. In particular, we find that this occurs for essentially the same 
parameter range that would make such 
objects detectable by Advanced LIGO-class detectors up to the Virgo Cluster. 
%This argument supports the idea that, 
%if magnetars are born fastly spinning, their GW emission could be detectable 
%by Advanced LIGO-class detectors up to the Virgo Cluster.
%%%%%%
%%%%%%
If our argument holds for at least a fraction of newly formed magnetars, then 
these objects constitute a promising new class of gravitational wave emitters. 

\keywords{gravitational waves --- stars: magnetic fields ---  stars: neutron 
--- stars: individual: SGR 1806-20)}
\PACS{97.60.Jd \and 97.60.Bw \and 04.30.Db \and 95.85.Sz}
%\PACS{ 97.60.Jd \and 97.60.Bw \and }
\end{abstract}

\section{Introduction}
\label{intro}
The Soft Gamma  Repeaters, SGRs, and the Anomalous X-ray Pulsars, AXPs, have 
a number of properties in common \citep{MeSte95,Kou98,WooTho04}. 
They have spin periods of  $\sim 5\div10$~s, spin-down 
secularly with $\sim 10^4\div10^5$~yr timescale, 
are isolated and in some cases associated to 
supernova remnants with $\sim 10^3\div10^4$~yr ages. Rotational energy losses 
are $10 \div 100$ times too low to explain the $\sim 10^{34}\div10^{35}$ erg/s 
persistent emission of these sources.  Both AXPs 
and SGRs have periods of intense activity during which recurrent, 
subsecond-long bursts are emitted (peak luminosities of $\sim 
10^{38} \div 10^{41}$~erg/s). 
The initial spikes of giant flares have comparable duration but 
3 to 6 orders of magnitude larger luminosity. Giant flares are rare, 
only three have been observed in about 30 yr of monitoring. Given the highly 
super-Eddington luminosities of recurrent bursts and, especially, giant flares,
accretion models are not viable.

In the magnetar model, SGRs and AXPs derive their emission from 
the release of the energy stored in their extremely high magnetic 
fields \citep{DT92,TD93,TD95,TD96,TD2001}. This is the leading 
model for interpreting the unique features of these sources.
According to it, a wound-up, mainly toroidal magnetic field 
characterizes  the neutron star interior ($B>10^{15}$~G).

The emerged (mainly poloidal) field makes up the neutron star 
magnetosphere; dipole strenghts ($B_d \sim $ few $\times 10^{14}$~G)
are required to generate the observed spin-down\citep{TD93,TM01}. 
Impulsive energy is fed to the neutron star 
magnetosphere through Alfv\'{e}n waves driven by local 
``crustquakes" and producing recurrent bursts with a large 
range of amplitudes. 
Giant flares likely originate in large-scale 
rearrangements of the toroidal inner field or catastrophic 
instabilities in the magnetosphere \citep{TD2001,Lyu03}.
Most of this energy breaks out of the magnetosphere in a 
fireball of plasma expanding at relativistic speeds which produces  
the initial spike of giant flares. The oscillating tail that follows this 
spike, displaying many tens of cycles at the neutron star spin, is 
interpreted as due to a ``trapped fireball", which remains anchored
inside the magnetosphere (the total energy released in this tail 
is $\sim 10^{44}$~erg in all three events detected so far, comparable to the
energy of a $\sim 10^{14}$ G trapping magnetospheric field).
\section{The 2004 December 27 Event and the Internal Magnetic field of 
Magnetars}
\label{sec:1}
The 2004 December 27 giant flare from SGR1806-20 provides a new estimate of the
internal field of magnetars. 
About $ 5 \times 10^{46}$~erg were released during the $\sim 0.6$~s long  
initial spike of this event \citep{Ter05,Hur05}. This is more 
than two decades higher than the energy of the other giant flares observed so 
far, the 1979 March 5 event from SGR 0526-66 \citep{Maz79} and the 1998 
August 27 event from SGR 1900+14 \citep{Hur99,Fer99}.
Only one such powerful flare has been recorded in about 30 yr of monitoring of 
the $\sim 5$ known magnetars in SGRs.  The recurrence time in a single 
magnetar implied by this event is thus about $\sim 150 $~yr. The 
realisation that powerful giant flares could be observed from distances of tens
of Mpc (and thus might represent a sizeable fraction of the short Gamma Ray 
Burst population) motivated searches for 2004 Dec 27-like events in the 
BATSE GRB database \citep{Laz05,Popo05}. The upper limits on the recurrence 
time of powerful giant flares obtained in these studies range from $\tau \sim 
130$ to 600~yr per galaxy, {\it i.e.} $\sim 4$ to 20 times longer than 
inferred above. Therefore these authors conclude that on 2004 Dec 27 we have 
witnessed a ``statistically unlikely'' event. 
On the other hand \citet{Tan05} find that the location of 10-25\% of the short 
GRBs in the BATSE catalogue correlates with the position of galaxies in the 
local universe ($< 110$~Mpc), suggesting that a fraction of the short GRBs 
may originate from a population of powerful Giant Flare-like events.

A 2004 Dec 27-like event in the Galaxy could not be missed, whereas several 
systematic effects can reduce the chances of detection from large distances 
(see e.g. the discussion in \cite{Laz05,Nak05}). Rather than 
regarding the 2004 Dec 27 event as statistically unlikely, 
one can thus evaluate the chances of a recurrence time of hundreds of years, 
{\it 
given} the occurrence of the 2004 Dec 27 hyperflare. We estimate that, having 
observed a powerful giant flare in our galaxy in $\sim 30$~yr of observations, 
the Bayesian probability that the galactic recurrence time is $\tau > 600$~yr 
is $\sim 10^{-3}$, whereas the $90\%$ confidence upper limit is $\tau \sim 
60$~yr. We thus favor smaller values and assume in the following $\tau \sim 
30$~yr. 

In $\sim 10^4$~yr (that we adopt for the SGR lifetime), about 70 very powerful 
giant flares should be emitted by an SGR, releasing a total energy of $\sim 4 
\times 10^{48}$~erg. 
We note that if the giant flares' emission were beamed in a fraction 
$b$ of the sky (and thus the energy released in individual flares a factor of 
$b$ lower), the recurrence time would be a factor of $b$ shorter. Therefore the
total release of energy would remain the same. If this energy originates 
from the magnetar's internal  magnetic field, this must be $\geq 10^{15.7}$~G
\citep{Ste05,Ter05}. This value should be regarded as a {\it lower limit}.   
Firstly, the magnetar model predicts a conspicuous neutrino luminosity from 
ambipolar diffusion-driven field decay, an energy component that is not 
available to flares. Including this, we estimate that the limit above 
increases by $\sim$ 60\% and becomes $B \geq 10^{15.9}$ G.
Secondly, ambipolar diffusion and magnetic dissipation should take place at a 
faster rate for higher values of the field \citep{TD96}. Therefore estimates 
of the internal B-field based on present day properties of SGRs 
likely underestimate the value of their initial magnetic field. 

Very strong toroidal B fields are expected to be generated inside a 
differentially rotating fast spinning neutron star, subject to vigorous 
neutrino-driven convection instants after its formation \citep{DT92}. 
A field of several $\times 10^{16}$ G can be generated in magnetars that 
are born with spin periods of a few milliseconds \citep{TD93}. As discussed 
by \citet{Dun98}, values up to $\sim 10^{17}$~G cannot be ruled out.

In the following we explore the consequences of these fields 
for the generation of gravitational waves from newborn magnetars. 
We parametrize their (internal) toroidal field with $B_{t,16.3} = B_t/
2\times 10^{16}$~G, (external) dipole field with $B_{d,14} = 
B_d/10^{14}$~G and initial spin period with $P_{i,2} = 
P_i/(2~\mbox{ms})$. 
\section{Magnetically-Induced Distortion and Gravitational Wave Emission}

The possibility that fast-rotating, magnetically-distorted neutron stars are
conspicuous sources of gravitational radiation has been discussed by several
authors \citep{Bona94,BoGo96}. More recently, work has been  
carried out in the context of the magnetar model, for internal magnetic fields 
strengths of $\sim 10^{14}-10^{16}$~G \citep{Kon00,Pal01,Cut02}.
In the following we show that for the range of 
magnetic fields discussed in Section 2, newly born, millisecond spinning 
magnetars are conspicuous sources of gravitational radiation that will be 
detectable up to Virgo cluster distances \citep{Ste05}. 

The anisotropic pressure from the toroidal B-field deforms a magnetar
into a prolate shape, with ellipticity $\epsilon_B \sim -6.4 \times
10^{-4} (<B^2_{t,16.3}>)$, where the brackets indicate a volume-average over 
the entire core \citep{Cut02}.
As long as the axis of the magnetic distortion is not aligned with the 
spin axis, the star's rotation will cause a periodic variation of the mass 
quadrupole moment, in turn resulting in the emission of gravitational waves, 
GWs, at twice the spin frequency of the star.\\ 
Free precession of the ellipsoidal NS is also excited and, as shown by 
\citet{MT72} and \citet{Cut02}, its viscous damping drives the symmetry axis 
of the magnetic distortion orthogonal to the spins axis, if the ellipsoid is 
prolate, \textit{i.e.} if the magnetic field is toroidal. 
Therefore, viscous damping of free precession in newly born magnetars leads to 
a geometry that maximizes the time-varying mass quadrupole moment, and GW 
emission accordingly.
%Therefore, viscous damping of 
%free precession has a shorter timescale than the NS spin-down, GWs will be 
%emitted with maximal efficienncy when the NS is still spinning at ms period.

However, the power emitted in GWs scales as $\propto P^{-6}$. Therefore the GW 
signal, for a given toroidal B-field, depends critically on the initial value 
and early evolution of the spin period. The spin evolution of a newborn 
magnetar is determined by angular momentum losses from GWs, electromagnetic 
dipole radiation and relativistic winds. According to \citet{ThChQu04},
the latter mechanism is negligible except for external dipole fields 
$< (6\div 7) \times 10^{14}$~G and we will neglect it here. 

The spin evolution of a newborn magnetar under the combined effects of GW and 
electromagnetic dipole radiation is given by 
 
\begin{equation}
\label{general}
\dot{\omega} = - K_d \omega^3 - K_{\mbox{\tiny{gw}}} \omega^5 \ ,
\end{equation}
where $\omega=2\pi/P$ is the angular velocity, \\
$K_d = (B^2_d R^6)/(6~I c^3)$ and $K_{\mbox{\tiny{gw}}} = (32/5) (G/c^5) I 
\epsilon^2_B $, with $R$ the neutron star radius, $G$ the gravitational 
constant and $c$ the speed of light. This gives a spin-down timescale of
\begin{equation}
\label{tausd}
\tau_{sd} \equiv \frac{\omega}{2 \dot{\omega}} 
\simeq 10~~P^2_{i,2}\left(B_{d,14}^2 + 1.15 B_{t,16.3}^{4}
P_{i,2}^{-2} \right)^{-1} \mbox{d} 
\end{equation}
The condition for the newly formed magnetar to become an orthogonal rotator 
before loosing a significant fraction of its initial spin energy is 
\citep{Ste05}:
\begin{equation}
\label{condition}
\frac{\tau_{sd}}{\tau_{ort}} \simeq 26~\frac{ B^2_{t,16.3}}{B^2_{d,14} P^{-1}_
{i,2} + 1.15~B^4_{t,16.3} P^{-3}_{i,2}} > 1 \ .
\end{equation}
If condition (\ref{condition}) is met, the magnetar quickly becomes a maximally
efficient GW emitter, while its spin period is still close to the initial one. 
In this case, the instantaneous signal strain can be expressed as: 
\begin{equation}
\label{strain}
h \sim 3 \times 10^{-26} d_{20}^{-1} P^{-2}_{2} B^2_{t,16.3}\ ,
\end{equation} 
where the distance $d_{20}=d/(20 \mbox{ Mpc})$  
is in units of the Virgo Cluster distance and the angle-averaged strain 
is that given by \citet{UshCutBil00}.      
We estimate the characteristic amplitude, $ h_{c} = h N^{1/2}$, 
where $N\simeq \tau_{sd}/P_i$ is the 
number of cycles over which the signal is observed. 
Using equation (\ref{strain}) we obtain:
\begin{equation} 
\label{effective}
h_{c} \simeq 6 \times 10^{-22} \frac{B_{t,16.3}^2} {d_{20}P^{3/2}_{i,2}
\left(B_{d,14}^2 + 1.15 B_{t,16.3}^4 P_{i,2}^{-2}\right)^{\frac{1}{2}}}
\end{equation}
Under the conditions discussed above, strong GW losses are not quenched 
immediately after the magnetar birth but rather extend in time, typically from 
days to a few weeks, before fading away as a result of the star spin-down. 
The characteristic amplitude in eq. (\ref{effective}) is within reach of 
GW interferometers of the Advanced LIGO class. 
\begin{figure*}
\centering
\includegraphics{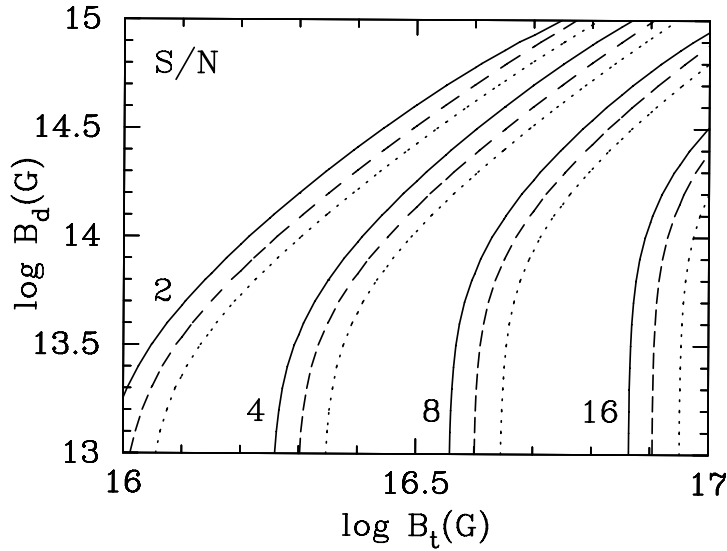}
%figure caption is below the figure
%\end{figure
%\begin{figure}
%\plotone{f1.eps}
\caption{Lines of constant $S/N$ for selected values of the 
initial spin period in the internal toroidal magnetic field, $B_t$, 
and external dipole field, $B_t$, plane for a source 
at the distance of the Virgo Cluster ($d_{20}=1$). Solid, dashed and dotted 
curves correspond to an initial spin period of $P_i=$1.2, 2 and 2.5~ms,
respectively. The calculations take into account the time required for 
the toroidal magnetic field axis to become orthogonal to the spin axis.  
Note that according to \citet{ThChQu04}, 
strong angular momentum losses by relativistic 
winds set in and dominate the spin down for $B_d > 6\div7 \times 10^{14}$~G;  
the curves for such values of $B_d$ should thus be treated with caution.}
\label{gwsall}
\end{figure*}
In order to assess the detectability of these GW signal we compute the optimal 
(matched-filter) signal-to-noise ratio for a signal sweeping the 
500 Hz - 2 kHz band by using the current baseline performance of Advanced 
LIGO (details are given in  \citet{Ste05}). Fig.~\ref{gwsall} shows lines of
constant $S/N$ for a source at $d_{20} = 1$  and selected values of the 
initial rotation period ($P_i = 1.2,2$ and 2.5~ms) in the $(B_t,B_d)$ plane: 
GWs from newborn magnetars can produce $S/N > 8$ for $B_t\ge 10^{16.5}$ G and 
$B_d \le 10^{14.5}$ G. This is the region of parameter space that offers
the best prospects for detection as we now discuss.
%%%%%%%%%
%%%%%%%%%
%\textbf{??? Simone il par qui sotto andrebbe scorciato e rivisto anche alla 
%luce delle tue cose piu' recenti con Virginia: pensaci tu ???}
%%%%%%%%%
%%%%%%%%%
Matched-filtering represents the optimal 
%(in the maximum likelihood sense) 
detection strategy for long-lived, periodic signals, also in those cases where
their frequency is slowly evolving over time. This process
%${\cal F}$-statistic \citep{Jara98}; its computation 
involves the correlation of the data stream with a discrete set of template
signals that probe the relevant space of unknown parameters: the sky position 
(2 \textit{extrinsic} parameters) and the 2 \textit{intrinsic} parameters 
that control the evolution of the GW phase (see Eq.(\ref{general})). Template
spacing in the parameter space is chosen appropriately, so as to 
%reduce the 
%maximum mismatch between the real signal and the best template to an 
%acceptably small value, \textit{i.e.} 
reduce to a value less than, say, 10\% (depending on the sensitivity one wants
to achieve) the fraction of the intrinsic signal-to-noise ratio that is lost 
in the cross-correlation. Dall'Osso \& Re (2006) have recently investigated in 
detail a matched-filtering (coherent) search strategy for the expected signal 
from newly formed magnetars in the Virgo cluster. Given the two (uncorrelated) 
parameters  involved in the signal frequency evolution, this approach implies 
an unaffordable computational cost. A huge number of templates $\sim 10^{18}$ 
would be required to cover the relevant parameter space (the plane
$B_d$ vs. $B_t$) with a sufficiently fine grid that the intrinsic 
signal-to-noise ratio would not be too degraded.
%A very conservative upper-limit on the required 
%number of filters involved in a search can be evaluated ignoring correlations 
%among the parameters: $\sim 100$ filters \citep{Brady98} are needed to cover 
%the $\sim 40\,\mathrm{deg}^2$ area of the Virgo cluster for a coherent 
%integration time $T\sim 10^6$~s, and $\sim (T\,f_i)^2 \sim 10^{18}$ filters 
%for the phase parameters, giving a total of $\sim 10^{20}$ templates. For
%this number of trials, we estimate that a signal with $S/N \sim 12$ 
%can be detected with  1\% false alarm and 10\% false dismissal 
%(if all the parameters of the signal were known the corresponding
%$S/N$ would be $\approx 4.6$). Of course the actual sensitivity limit of
%such a search is likely set by the available computational resources. In
%practice, the search for GWs from newborn magnetars is probably
%best carried out using a ``hierarchical search strategy'' where 
%coherent and incoherent stages are alternated in order to achieve 
%(quasi-)optimal sensitivity at affordable computational costs
%\citep{Brady00,Kri04,Cut05,Abb05}.
%Although the presence of a trigger based on the detection of the
%newborn magnetar by other means (e.g. the corresponding supernova)
%would reduce the region of the \textit{extrinsic} parameter space that needs 
%to be searched, these results are not affected sinc 
%One could rely on the increase in sensitivity that will be obtained with the 
%operation of the network of GW detectors that is coming on line. However, this
%may allow for just a factor of a few 
Although this calculation is rather idealized, the resulting number of 
templates is so large that no realistic calculation could decrease it by 
the several orders of magnitude needed to make the search feasible.
The search for this new class of signals represents however a challenge 
that must be investigated in greater depth. We are currently studying a 
hierarchical approach to the problem, a process where coherent and incoherent 
stages of the search are alternated as to reduce the computational 
requirements by a large factor, at the price of a relativeley modest loss in 
sensitivity to the signal. \\
%the week-long signal is split in several 
%shorter ``stacks'' that are summed in phase and the resulting short data
%time-series 
%%%%%%%%%%%
%%%%%%%%%%%
\section{Observational Constraints on GW emission from newly formed magnetars}
%\textbf{??? Simone metti qui le cose del tuo talk ???}
%Recent X-ray observations made with XMM-Newton of SNRs around magnetar 
%candidates have provided for the first time an observational basis to discuss 
%possible formation scenarios of these objects (Vink et al. 2006). \\
%We discuss here a straightforward application of our model of GW emission 
%from newly formed magnetars to the results of these observations. Despite 
%uncertainties being yet quite large, some general conclusions of particular 
%interest can already be drawn. The subject is quite new and prospects for 
%future progress are indeed encouraging.\\
A NS spinning at $\sim$ ms period has a spin energy $E_{spin} \approx 2.8 
\times 10^{52} (P_i / 1~\mbox{ms})^{-2}$ erg and the spindown timescale 
through magnetic dipole radiation and/or relativistic particle winds is 
extremely short, from a few weeks to $\sim$ one day for dipole fields in
the ($10^{14} \div 10^{15})$ G range. In standard magnetar scenarios, most of 
the initial spin energy is expected to be rapidly transferred to the 
surrounding supernova ejecta through these spindown mechanisms 
\citep{ThChQu04}.
Therefore, present-day SNRs around known magnetar candidates should bear the 
signature of such a large energy injection. For initial spin periods less than 
3 ms, the injected energy would be $> 3.5 \times 10^{51}$ erg, making these 
remnants significantly more energetic than those surrounding ordinary NSs 
($\leq 10^{51}$ erg s$^{-1}$).\\
The X-ray spectra of the SNRs surrounding known magnetar candidates (two APXs 
and two SGRs) studied by Vink et al. (2006), do not show any evidence that 
their total energy content differs from that in remnants surrounding common 
NSs ($\approx 10^{51}$ erg): this result constrains magnetar parameters at 
birth. 
%However, constraints depend significantly on the very model assumed. 
\citet{Vink} deduced from their measurements an initial 
spin $P_i \geq (5\div 6)$ ms for the above mentioned sources, assuming that 
all the spin-down energy is emitted through electromagnetic radiation 
and/or particle winds, and thus absorbed by the surrounding ejecta in the 
early days of spindown. Their limit period is long enough to rise a serious 
question as to the viability of the $\alpha-\Omega$ dynamo scenario for 
generating the large-scale magnetic fields of magnetars. Models that do not 
rely upon very short spin periods at birth, such as the flux-freezing scenario 
suggested by \citet{FeW06}, would be favored by these results.\\
%based on the magnetar model developed 
%by Thomspon \& Duncan (1995, 1996, 2001) and on the recent work by Stella et 
%al. (2005), to show tha
%Our model for GW emission from newly  formed magnetars, however, can be used 
%to pursue further a quite different possibility.
%to constrain a combination 
%of initial 
%spin period ($P_i$), internal ($B_t$) and external ($B_d$) magnetic field of 
%magnetars at birth. 
However, given the possibility that newly formed magnetars be strong GW 
emitters in their early days, we show that the results by \citet{Vink} can be 
accounted for within this framework. Most of a magnetar's 
initial spin energy could indeed be released through GWs, without being 
absorbed by the expanding remnant shell.
%assuming that these objects were formed fastly 
%spinning, with very strong internal fields (in the $10^{16}$ G range) and 
%releasing a significant fraction of their spin energy through gravitational 
%waves. 
%Interestingly, our conclusion is very close to that of Arons (2003), 
%whcih was based on a different and independent argument. Finally we note 
%that, recently, Kaminker et al. (2006) have reached the conclusion that if 
%magnetars thermal emission is indeed powered by magnetic-field-induced 
%internal heating,
%then the thermal conduction properties of magnetized NS matter require the
%internal field to be $> 10^{16}$ G. Remarkably, this is veru close to the
%absolute lower limit derived by Stella et al. (2005) based on the energetics 
%and likely recurrence time of Hyper Flares such as the Dec 27 event.\\
%
%
%%%Therefore, observations of present-day SNRs around \\AXPs/SGRs can be 
%%%interpreted in two ways: either magnetars do not form with the very fast 
%%%spin required for the $\alpha-\Omega$ dynamo to generate extremely large 
%%%internal fields or, if they form with such fast spins, they most likely 
%%%have internal magnetic fields in excess of $10^{16}$ G and spin down mostly 
%%%through the emission of GWs. 
Therefore, the results of X-ray studies can be used within our model to 
constrain the initial combination of $P_i, B_d, B_t$.
%Besides suggesting a very 
%interesting would make them likely sources 
%for next generation GW detectors, such as LIGO II, even at the distance of 
%the Virgo Cluster, as discussed recently by Stella et al. (2005).
\subsection{Strong GW emission at birth?}
\label{thecase}
The general expression for the total energy emitted via GWs is given by:
%: this is the fraction of energy, with respect to the total, that is 
%emitted through GWs at each spin frequency, integrated over the whole spin 
%history.
%
\begin{equation}
\label{integrategw}
E^{\mbox{\tiny{TOT}}}_{\mbox{\tiny{gw}}} = - \int_{t_i}^{\infty} \dot{E}_{\mbox{\tiny{gw}}}~dt = -
\int_{\omega_i}^0 \frac{\dot{E}_{\mbox{\tiny{gw}}}}{\dot{\omega}}d\omega
\end{equation}
%
%where the spin-down model of Eq. (\ref{general}), 
%Note that
%The last step accounts for the spin-down being driven by magnetic dipole 
%radiation, as well as GWs, so that 
%this the fraction of energy, with respect to the total, that is emitted 
%through GWs at each spin frequency, integrated over the whole spin history.
%
\begin{figure*}
\centering
\includegraphics[angle=0, width=3.5in]{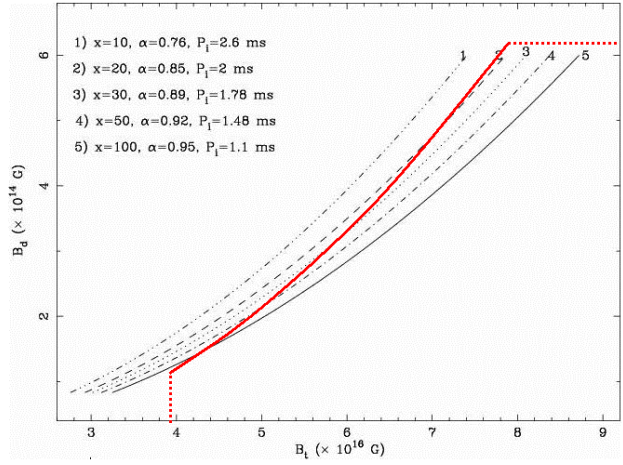}
%figure caption is below the figure
%\end{figure
%\begin{figure}
%\plotone{f1.eps}
\caption{Loci in the $B_d$ vs. $B_t$ plane for which 
$E^{\mbox{\tiny{inj}}}_{SN} \simeq 10^{51}$ ergs. Curves are labelled for 
different values of the ratio ($x$) of the GW to magnetodipole torques. 
Through Eq. (\ref{secondrelation}), it is seen that fixing this ratio is 
equivalent to defining the corresponding initial spin. The corresponding locus 
is thus uniquely determined through the first expression in Eq. (\ref{defX}).
Note that, given $x$, 
the corresponding curve shifts to the right for decreasing values of 
$E^{\mbox{\tiny{inj}}}_{SNR}$. The red curve represents the S/N=8 curve of 
Fig. (\ref{gwsall}).
In the framework of our model, thus, fast spinning newly formed magnetars that 
do emit $\leq 10^{51}$ ergs of spin-down energy 
through electromagnetic radiation can be expected to be detectable with LIGO 
II up to the distance of the Virgo cluster. Stated the other way, objects 
whose GW signal could be detectable by LIGO II up to the distance of the Virgo 
Cluster would inject - as found in the case of magnetar candidates in the 
Galaxy - $\leq 10^{51}$ ergs in the expanding shells of the surrounding SNR.}
\label{Bplane}
\end{figure*}
%
%With the spin-down law
%
%\begin{equation}
%\label{pdot}
%\dot{\omega} = -~\underbrace{\frac{B^2_d R^6}{6I c^3}}_{K_d/I}~\omega^3 -~
%\underbrace{\frac{32}{5}\frac{GI\epsilon^2_B}{c^5}}_{K_{gw}/I}~\omega^5
%\end{equation}
%
We insert Eq. (\ref{general}) and the expression for the GW luminosity of an 
elliptically distorted, spinning object into Eq. (\ref{integrategw}) to 
obtain the following analytical solution:
\begin{equation}
\label{solveintegral}
E^{\mbox{\tiny{TOT}}}_{\mbox{\tiny{gw}}} = I \int_{0}^{\omega_i} 
\frac{\omega^3}{\omega^2 + A}d\omega = I\left[\frac{\omega^2}{2} - 
\frac{A}{2}\mbox{ln}(\omega^2+A)\right]^{\omega_i}_0 
\end{equation}
where $A = K_d/K_{\mbox{\tiny{gw}}}$. Since $E^{\mbox{\tiny{TOT}}}_{\mbox{\tiny{gw}}}$ 
amounts to a fraction $\alpha$ of the initial spin energy of the NS, we can 
write:
\begin{equation}
\label{related}
(1-\alpha) \omega^2_i = A~\mbox{ln} \left(\frac{\omega^2_i+A}
{A}\right)
\end{equation}
Finally, defining the ratio of the GW over magnetodipole torque as $x \equiv 
(\omega^2_i/A)$:
\begin{eqnarray}
\label{defX}
x & \approx & 2.25~\frac{B^4_{t,16.3}}{B^2_{d,14}~P^2_{i,2}}\nonumber
\\
1-\alpha & = & \frac{\mbox{ln}(1+x)}{x}
\end{eqnarray} 
The above expressions provide us with two relations between the five
parameters ($\omega_i, B_d, B_t, x, \alpha$). A third one derives from the 
fact that the remaining fraction ($1-\alpha$) of the initial spin energy is 
available for being transferred to the ejecta through magnetodipole radiation. 
By assuming that all this energy is effectively transferred to the SNR, we get:
\begin{equation}
\label{secondrelation}
\omega^2_i = \frac{2~E^{\mbox{\tiny{inj}}}_{SNR}}{(1-\alpha) I} =
\frac{2x~E^{\mbox{\tiny{inj}}}_{SNR}}{\mbox{ln}(1+x) I}
\end{equation}
where $E^{\mbox{\tiny{inj}}}_{SNR}$ is the spin-down energy injected in the 
expanding shell, constrained by the results of \citet{Vink} to be 
$\simeq 10^{51}$ erg.
%
%\subsection{NS parameters and dependence of results on the NS radius}
%\label{parameters}
%In the next subsection we will try to account for some, even 
%small, fraction of the resiidual $(1-\alpha)$ initial energy being lost to 
%other channels or, ultimately, being not transferred to the remnant.\\
We use the results of \citet{LaPr}, according to which a 1.4 M$_{\odot}$ NS 
has a radius $R \simeq 12$ km and a moment of inertia $I \approx 0.35 M R^2 
\simeq 1.4 \times 10^{45}$ g cm$^2$. Note that, by these numbers, one 
obtains the following relation between the measured $P$ and $\dot{P}$ of NSs 
and the corresponding value of the dipolar magnetic field at the magnetic pole 
(our $B_d$)\footnote{This 
is a factor 1.5 less than usually assumed with $I = 10^{45}$ g cm$^2$ and 
$R=10$ Km. Use of the most up to date parameters is required, given the strong 
dependence of the two competing torques on the exact value of the magnetic 
fields.}:
\begin{equation}
\label{dipolefield}
B_d \simeq 4.4 \times 10^{19}~(P\dot{P})^{\frac{1}{2}}~\mbox{G}
\end{equation}
%
%This gives values of in the interval $(0.8\div 6) \times 10^{14}$ G for six 
%out of seven AXPs and in the interval $(7.5\div 10) \times 10^{14}$ G for 
%SGRs (cfr. \citep{WooTho04}). However, while AXPs are, in general, reasonably 
%stable rotators whose spin-down can be related to the intensity of their 
%dipolar field, the SGRs are known to have strongly variable period 
%derivatives and, in general, a very noisy spin-down. This fact suggests that 
%SGRs spin-down is likely to be significantly affected by additional 
%mechanisms, apart from magnetodipole radiation, and that their inferred 
%dipole fields may be subject to large uncertainties.\\
As can be seen from Fig. (\ref{gwsall}), dipole magnetic fields stronger than 
$(5\div6) \times 10^{14}$ G rapidly quench the expected S/N ratio of GW 
signals from newly formed magnetars. Therefore, we restrict our investigation 
to polar dipole fields $B_d < 6$.
%, since for larger fields there are no 
%realistic prospects for significant GW signals to be emitted. We note that 
%this range includes the inferred magnetic fields of AXPs but not those of 
%SGRs.
%The latter are, however, likely affected by additional mechanisms apart from 
%magnetodipole radiation, and their inferred dipole fields may be subject to 
%larger uncertainties than those of AXPs \citep{WooTho04}.\\
%As a last step, we write down the explicit expression for $x$, by normalizing 
%the physical parameters as usual:
%Normalizing the physical parameters as $P_{i,\mbox{\small{ms}}} = 
%(P_i/1~\mbox{\tiny{ms}})$, $B_{d,14} = (B_d/10^{14}~\mbox{G})$ and $B_{t,16.3}
%= (B_d/2\times 10^{16}~ \mbox{G})$, $x$ is simply expressed as:
%
%\begin{equation}
%\label{xsimple}
%x \approx 2.25~\frac{B^4_{t,16.3}}{B^2_{d,14}~P^2_{i,2}}
%\end{equation}
%
%Therefore, given $x$ one obtains $\omega_i$ and, eventually, the appropriate 
%curve in the $B_d$ vs. $B_t$ plane.\\
%Steps 1) to 3) are shown in Figs. \ref{icsalfa}, \ref{omegainiz} and 
%\ref{BdBt}, respectively. In the last figure, four curves in the $B_d$ vs.
%$B_t$ plane have been drawn for five different values of $x = 1, 10, 20, 30$
%and $50$.
%
%We are now in the position to use results of these observations to put 
%self-consistent constraints on the physical parameters at birth of those 
%magnetars, within the framework discussed here.
From Eq. (\ref{secondrelation}) we see that, once $x$ is given, both $\alpha$ 
and $\omega_i$ are determined (one can indeed use any of the three as the free
parameter). The first expression of Eq. (\ref{defX}) thus provides a relation 
between $B_d$ and $B_t$.\\
We have repeated this procedure for five values of the ratio of the GW over 
the magnetodipole torque ($x$)
%From the last expression in Eq. (\ref{secondrelation}) we can find a very 
%simple expression for the ratio $\omega^2_i / x = A$, so that
%This will be done in two
%steps:\\
%
%(i) insert the relation $\alpha(x)$ into eq. (\ref{secondrelation}), together
%    with the measured value of $E^{\mbox{\tiny{inj}}}_{SNR}$, in order to
%    obtain $\omega_i$ as a function of $x$, from which 
%\begin{equation}
%\label{useful}
%\omega^2_i = \frac{2x~E^{\mbox{\tiny{inj}}}_{SNR}}{\mbox{ln}(1+x) I}
%\end{equation}
%
%Given the definition 
%(ii) insert $\omega_i(x)$ (eq. \ref{useful}) in the definition of $x$, that
%gives:
%
%\begin{equation}
%\label{final}
%A = \frac{2~E^{\mbox{\tiny{inj}}}_{SNR}}{I \mbox{ln}(1+x)}
%\end{equation}
%
and, for each of them, obtained the implied values of $\omega_i$ and $\alpha$ 
and a curve $B_t$ vs. $B_d$. \\
%
%Therefore, given $x$ one obtains $\omega_i$ and, eventually, the appropriate 
%curve in the $B_d$ vs. $B_t$ plane.\\
Fig. \ref{Bplane} summarizes our results. Loci in the $B_d$ vs. $B_t$ plane 
for which fast spinning, ultramagnetized magnetars are consistent with the 
energetic constraints derived by \citet{Vink} are drawn for the five 
chosen values of $x$ or, equivalently, the initial spin period (since 
$E^{inj}_{SN}$ is fixed). Details are given in the caption. \\
In summary, we have first identified a range of initial conditions (spin 
period, internal and external magnetic field), within which newly formed 
magnetars can be interesting targets for next generation GW detectors. Then 
we have calculated that, within most of that same region, magnetars should 
emit less than $10^{51}$ ergs through magnetodipole radiation (cfr. 
\citet{Dalet} and Fig. \ref{Bplane}). This is compatible with the
limits inferred through recent X-ray observations of SNRs around present-day 
magnetar candidates. \\
%appears to have absorbed less than $10^{51}$ erg from the central, spinning 
%down NS: the gas bears no signature of the expected spin energy of the 
%magnetar when it first formed ($> 3\times 10^{51}$ erg).\\
%These results can be interpreted in two simple ways: either magnetars do not 
%form with the very fast spin required for the $\alpha-\Omega$ dynamo to 
%operate, their initial spin energy needs not being very high and their 
%potential GW emission is not peculiar or, if they do form with fast spins, 
%their internal magnetic fields should exceed $10^{16}$ G and they should spin 
%down mostly through the emission of GWs, in the first days from their 
%formation. In this case, they have all the potential to become interesting 
%targets for the next generation of GW detectors, being observable up to the 
%Virgo cluster distance. 
%naturally explained if these objects were indeed formed with a very fast spin
%and a very high internal magnetic field. They could have radiated away most of
%their spin energy through GWs, rather then electromagnetically.
%
%%%%%%%%%%%
%%%%%%%%%%%
\section{Discussion}
The energy liberated in the 2004 December 27 flare from SGR~1806-20, together 
with the likely recurrence rate of these events, points to a magnetar internal 
field strength of $\sim 10^{16}$~G or greater.
Such a field likely results from differential rotation in a millisecond 
spinning proto-magnetar and deforms the star into a prolate shape. 
Magnetars with these characteristics are expected to be very powerful 
sources of gravitational radiation in the first days to weeks of their life. 
An evolving periodic GW signal at $\sim 1$~kHz, whose frequency halves 
over weeks, would unambiguously reveal the early days of a fast spinning 
magnetar.

Prospects for revealing their GW signal depend on the birth rate 
of these objects. 
The three associations between an AXP and a supernova remnant 
(ages in the $10^3 \div 10^4$~yr range) implies a magnetar birth rate of 
$\geq 0.5\times 10^{-3}$~yr$^{-1}$ in the Galaxy \citep{Gae99}. 
Therefore the chances of witnessing the formation of a magnetar in our Galaxy 
are slim. A rich cluster like Virgo, containing $\sim 2000$ galaxies, 
is expected to give birth to magnetars at a rate of $\geq$ 1 yr$^{-1}$. 
A fraction of these might have sufficiently high toroidal fields that a 
detectable GW is produced. \\
It has been recently found that SNR shells around some magnetar candidates 
have comparable expansion energies to standard SNR shells. This implies that 
either the NS was not initially spinning as fast as required for an 
$\alpha-\Omega$ dynamo to amplify its field to magnetar strengths, or that 
most of its initial spin energy was emitted in a way that did not interact
with the ejecta. GWs have indeed such property. If the internal magnetic 
field of newly formed magnetars is $> 10^{16}$ G, comparable to the lower 
limit estimated through the enegetics of the Dec 27 Giant Flare, then their
GW emission can be strong enough to radiate away most of their initial spin
energy. The required amount of energy emitted through GWs is indeed such that, 
had these magnetar candidates been at the distance of the Virgo cluster, they 
would have been revealed by a LIGO II-class GW detector.

Therefore, GWs from newly formed magnetars can account naturally for the 
recent X-ray observations of SNRs around galactic magnetars. The main 
conclusion that can be drawn at present is that newborn, fast spinning 
magnetars represent a potential class of GW emitters over Virgo scale 
distances that might well be within reach for the forthcoming generation of 
GW detectors.
\end{document}